\documentclass{aa}  

\usepackage{graphicx}
\usepackage{txfonts}
\usepackage{natbib}
\bibpunct[; ]{(}{)}{;}{a}{}{,}
\usepackage{multirow}
\usepackage{hyperref}
\usepackage{color}

\begin{document} 

\title{Ground-based transit observations of the super-Earth
  GJ\,1214\,b\thanks{Based on observations obtained at the Southern
    Astrophysical Research (SOAR) telescope, which is a joint project
    of the Minist\'{e}rio da Ci\^{e}ncia, Tecnologia, e
    Inova\c{c}\~{a}o (MCTI) da Rep\'{u}blica Federativa do Brasil, the
    U.S. National Optical Astronomy Observatory (NOAO), the University
    of North Carolina at Chapel Hill (UNC), and Michigan State
    University (MSU). SofI results based on observations made with ESO
    Telescopes at the La Silla Paranal Observatory under programme ID
    087.C-0509.}}
\author{
C. C\'aceres\inst{1,2}%
\and%
P. Kabath\inst{3}\fnmsep%
\and%
S. Hoyer\inst{4,5}\fnmsep%
\and%
V. D. Ivanov\inst{3}\fnmsep%
\and%
P. Rojo\inst{6}\fnmsep%
\and%
J. H. Girard\inst{3}\fnmsep%
\and%
E. Miller-Ricci Kempton\inst{7}\fnmsep%
\and%
J. J. Fortney\inst{8}\fnmsep%
\and%
D. Minniti \inst{9,10}\fnmsep%
}

\institute{%
  Instituto de F\'{\i}sica y Astronom\'{\i}a, Universidad de
  Valpara\'{\i}so, Av. Gran Breta\~{n}a 1111, Valpara\'{i}so 2360102, Chile\\
  \email{ccaceres@dfa.uv.cl}%
  \and
  Millenium Nucleus ``Protoplanetary Disks in ALMA Early Science'', Universidad de Valpara\'{\i}so, Valpara\'{\i}so 2360102, Chile%
  \and%
  European Southern Observatory, Av. Alonso de C\'{o}rdova 3107,
  Vitacura, Santiago 19001, Chile%
  \and%
  Instituto de Astrof\'{\i}sica de Canarias, V\'{i}a L\'{a}ctea s/n,
  E38200 - La Laguna, Tenerife, Canary Islands, Spain.%
  \and
  Department of Astrophysics, University of La Laguna, V\'{i}a L\'{a}ctea s/n,
  E38200-La Laguna, Tenerife, Canary Islands, Spain.%
  \and%
  Universidad de Chile, Camino El Observatorio 1515, Las Condes,
  Santiago, Chile%
  \and%
  Department of Physics, Grinnell College, Grinnell, IA 50112, USA%
  \and%
  University of California, Santa Cruz, Department of Astronomy \&
  Astrophysics, 1156 High Street, Santa Cruz, CA 95064, USA%
  \and%
  Pontificia Universidad Cat\'olica de Chile, Departamento de
  Astronom\'{\i}a y Astrof\'{\i}sica, Av. Vicu\~na Mackenna 4860,
  Macul, Santiago, Chile.%
  \and%
  Specola Vaticana, V00120 Vatican City State, Italy.%
 }
   \date{Received September 15, 1996; accepted March 16, 1997}

 
\abstract
{GJ\,1214\,b is one of the few known transiting super-Earth-sized
  exoplanets with a measured mass and radius. It orbits an M-dwarf,
  only 14.55\,pc away, making it a favorable candidate for follow-up
  studies. However, the composition of GJ\,1214\,b's mysterious
  atmosphere has yet to be fully unveiled.}
{Our goal is to distinguish between the various proposed atmospheric
  models to explain the properties of GJ\,1214\,b: hydrogen-rich or
  hydrogen-He mix, or a heavy molecular weight atmosphere with
  reflecting high clouds, as latest studies have suggested.}
{Wavelength-dependent planetary radii measurements from the transit
  depths in the optical/NIR are the best tool to investigate the
  atmosphere of GJ\,1214\,b. We present here (i) photometric transit
  observations with a narrow-band filter centered on 2.14\,$\mu$m and
  a broad-band $I$-Bessel filter centered on 0.8665\,$\mu$m, and (ii)
  transmission spectroscopy in the $H$ and $K$ atmospheric windows
that   cover three transits. The obtained photometric and
  spectrophotometric time series were analyzed with MCMC simulations
  to measure the planetary radii as a function of wavelength. We
  determined radii ratios of $0.1173_{-0.0024}^{+0.0022}$ for $I$-Bessel
  and $0.11735_{-0.00076}^{+0.00072}$ at 2.14\,$\mu m$.}
{Our measurements indicate a flat transmission spectrum, in agreement
  with last atmospheric models that favor featureless spectra with
  clouds and high molecular weight compositions.}
{}

\keywords{
planetary systems, 
techniques: photometric, 
techniques: spectroscopic, 
planets and satellites: atmospheres,
stars: individual: GJ 1214
}

\maketitle
%

\section{Introduction}

The handful of known transiting extrasolar super-Earths have been
mostly found by space missions such as {\it Kepler}, {\it CoRoT}, or {\it
  MOST}, with the notable exception of GJ\,1214\,b (2.7\,$R_{\oplus}$,
6.55\,$M_{\oplus}$), which was discovered by the ground-based
transiting survey {\it MEarth} around a bright nearby M-star
\citep{charbonneau_etal2009}. Despite the relatively small radius of
GJ\,1214\,b, the small stellar radius of its host star results in a
$\sim$1.5\% transit depth, making it one of the super-Earths best
suited for follow-up studies.

The recent theoretical predictions for the atmosphere of GJ\,1214\,b
currently offer the most plausible models that fit the planet's mass,
radius, and irradiation level: a rocky or an icy core with a nebular
hydrogen-helium envelope, that is, a mini-Neptune, a rocky planet with
an outgassed hydrogen atmosphere, or a core with a heavy (up to 45\%
of the planet mass) hot water vapor envelope with high molecular mass
\citep{rogers_seager_2010, nettelmann_etal2011}, as well as a planet
with a cloudy or hazy atmosphere, with a high mean molecular mass
composition \citep{morley_etal2013}.

GJ\,1214\,b has been the target of an intensive observing campaign.
\citet{bean_etal2010} reported a flat optical transmission spectrum
that ruled out a cloud-free mini-Neptune model. This conclusion was
supported by additional transit observations of \citet{bean_etal2011},
\citet{berta_etal_2012}, and \citet{crossfield_etal_2011} in the
near-infrared (NIR) region, and \citet{desert_etal2011} in the
mid-infrared. However, \citet{croll_etal2011} measured a deeper
2.2\,$\mu$m transit, implying H$_2$ absorption, consistent with a
mini-Neptune model. Transient achromatic haze might reconcile these
observations, but the close dates of the $K$-band observations of
\citet{croll_etal2011} and \citet{crossfield_etal_2011} make it
unlikely, and as \citet{berta_etal_2012} pointed out, there is no
known source of achromatic haze. Strangely, \citet{narita_etal2012}
reported $J-$ and $H-$band observations consistent with a flat
featureless transmission spectrum, but shallower $K_S$-band
transmission depth, in disagreement with the result of
\citet{croll_etal2011}. \citet{fraine_etal_2012} also reported
constant planetary atmosphere radii at $I+z$, 3.6\,$\mu$m, and
4.5\,$\mu$m. Finally, \citet{murgas_etal2012} found a radius of
GJ\,1214\,b at the H$\alpha$ line higher than the radii measured
at nearby continua, but the difference is not statistically
significant. The latest observations reported by
  \citet{teske_etal2013}, \citet{colon_gaidos2013},
  \citet{wilson_etal2013}, \citet{gillon_etal2013},
  \citet{demooij_etal2013}, and \citet{kreidberg_etal2014} have agreed
  in showing a featureless spectrum, favoring high mean molecular
  mass compositions. In that context, recently \citet{kreidberg_etal2014} ruled out the cloud-free high
  molecular weight atmosphere scenario for GJ\,1214b with a high
  statistical significance.

Here we report new optical and NIR ground-based observations of
GJ\,1214\,b transits, aiming to independently verify these results.
The paper is organized as follows: Sect.\,\ref{sec:obs} presents our
observations, Sect.\,\ref{sec:analysis} describes the analysis,
Sect.\,\ref{sec:disc} presents a discussion of our results, and
Sect.\,\ref{sec:summary} is a summary of the main points of this paper.

\section{Observations and data reduction\label{sec:obs}}

\begin{figure}[!t]
\begin{center}
\includegraphics[width=0.48\textwidth]{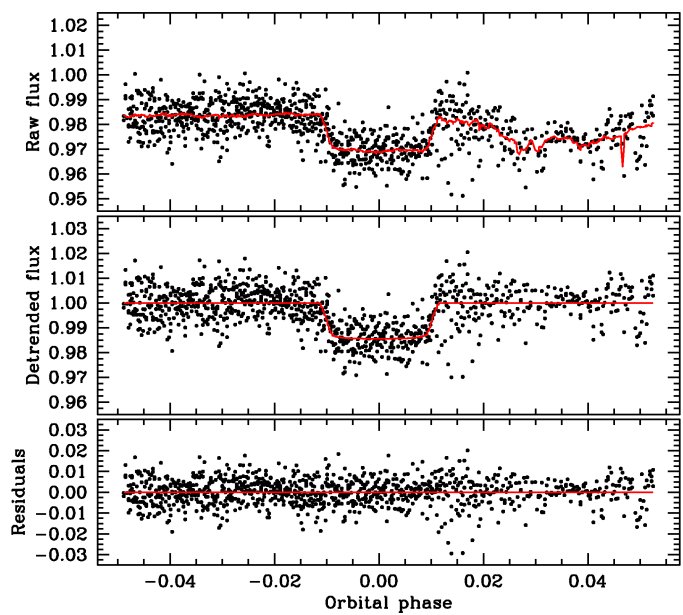}
\end{center}
\caption{\label{fig:osirislc} $2.14\,\mu m$ GJ\,1214\,b
  photometric series obtained with OSIRIS. {\sl Top:} Raw light curve,
  with the best-fitting transit model in red. {\sl Middle:} De-trended
  light curve with the best-fitting model. {\sl Bottom:} Residuals
  from the best-fitting model; the RMS. is 0.0062. A color version
  of this plot can be found in the electronic version of the
  paper.}%
\end{figure}

\begin{figure}[!t]
\begin{center}
\includegraphics[width=0.49\textwidth]{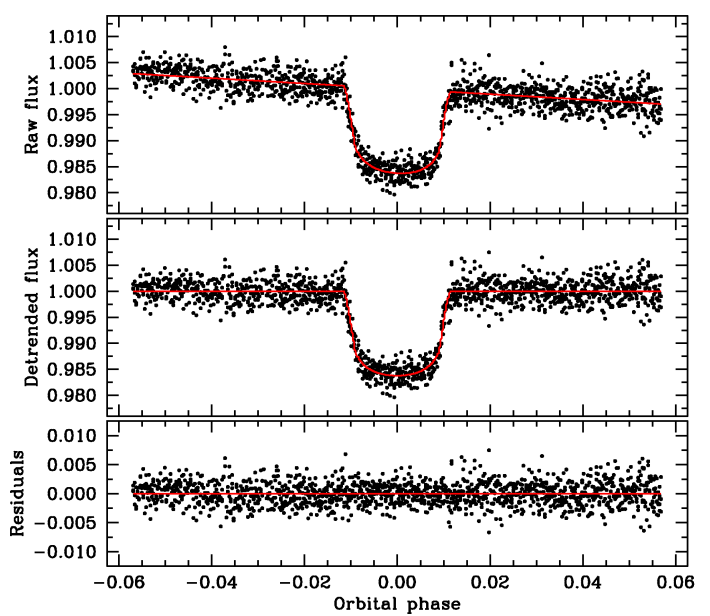}
\end{center}
\caption{\label{fig:soilc}  $I$-Bessel GJ\,1214\,b photometric
  series obtained with SOI. {\sl Top:} Raw light curve, with the best-fitting transit model in red. {\sl Middle:} De-trended light curve
  with the best-fitting model. {\sl Bottom:} Residuals from the best-fitting model; the RMS is 0.0019. A color version of this plot
  can be found in the electronic version of the paper.}
\end{figure}

\subsection{NIR photometry with { \it OSIRIS} }

We observed a GJ\,1214\,b transit from UT 0:19 to 4:10 on the night of
August 09, 2011 with the {\it Ohio State Infrared Imager/Spectrometer}
\citep[{\it OSIRIS};][]{depoy_etal_1993} at the 4.1 m Southern
Astrophysical Research ({\it SOAR}) Telescope at Cerro Pach\'on,
Chile.  {\it OSIRIS} only uses a 577$\times$577\,px window from a
1024$\times$1024 px HgCdTe array, yielding 191$\times$191\,arcsec FoV
(pixel scale 0.331\,arcsec\,px$^{-1}$). A narrow-band ($1$\%) filter
centered on 2.14 $\mu$m was used. We placed GJ\,1214 and a nearby
reference star (2MASS\,J17152424+0455041) within the FOV. They have
similar apparent brightness and are separated by $\sim$3\,arcmin.

The observations were carried out in stare mode, that is, the stars were
kept on the same positions to minimize the systematic effects
associated with inaccurate flat-fielding and intra-pixel sensitivity
variations. A total of 1,030 science images were collected, with
exposure times ranging from 5 to 20\,seconds, which had to be adjusted
to avoid reaching the nonlinearity regime under variable seeing and
airmass.

First, we subtracted the dark current by scaling one series of darks to
match the different exposure times used throughout the night, and
flat-fielded the images. To extract the apparent fluxes from the
reduced images we carried out aperture photometry with the IRAF {\it
  Digiphot} package\footnote{IRAF is distributed by the NOAO, which
  are operated by the AURA in Astronomy, Inc., under cooperative
  agreement with the NSF.}. Led by our previous experience with
stare-mode NIR data, we removed the sky by measuring it in circular
annuli, contiguous to the aperture centered on the object
\citep{caceres_etal2009,caceres_etal2011}. We performed photometry for
different aperture radii, ranging from 4.0 to 14.0\,px, in steps of
0.5\,px, and selected the combination of source and sky apertures
that minimizes the RMS of the final differential light curve:
target and reference aperture radii of 2.32\,arcsec, and a sky annulus
with 4.9\,arcsec inner radius and 3.3\,arcsec width.  The generated
light curve is shown in Fig.\,\ref{fig:osirislc} (top).

\subsection{Optical photometry with {\it SOI}}

A transit of GJ\,1214\,b was observed in the $I$-Bessel filter from UT
04:03 to 09:30 on the night of April 28, 2010 with the SOAR Optical
Imager ({\it SOI}) at the {\it SOAR} telescope on Cerro Pach\'on. The
instrument has a mosaic of two E2V 2$\times$4\,k CCDs, covering a
$\sim$5.5$\times$5.5\,arcmin FoV (pixel scale
0.077\,arcsec\,px$^{-1}$).  The binning was 2$\times$2, and the
readout time was 11\,seconds.

We placed the target close to the center of the FoV, which allowed us
to observe eight reference stars of similar or slightly fainter
brightness to that of the target. We obtained 1,730 frames, with
integration times between 3 and 5\,seconds--adjusted throughout the
night to keep the core of the stellar images in the linear regimen.

The basic data reduction included trimming the images, bias
subtraction and flat-fielding corrections with SOI's custom-made
pipeline \citep{hoyer_etal2012}. Next, we performed standard
aperture photometry of the target and the eight reference stars. The
optimal aperture and sky-annulus radii were selected to minimize the
RMS of the portions of the light curve before and after the
transit, best values being stellar aperture radius - 18\,pix, sky
annulus radius - 20\,pix and 10\,pix in width.  Similarly, our experiments
with the reference stars demonstrated that the best light curve is
obtained using only one of them -- the brightest
2MASS\,J17152424+0455041 ($K_S$=8.983\,mag). According to our tests,
adding more reference stars increased the noise. The generated light
curve is shown in Fig.\,\ref{fig:soilc} (top).

\begin{figure}[!t]
\begin{center}
\includegraphics[width=84mm]{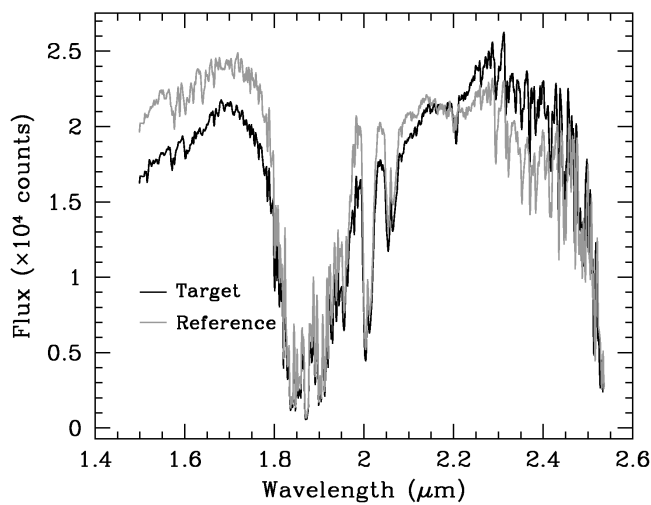}
\end{center}
\caption{Typical wavelength-calibrated one-dimensional spectrum of 
GJ\,1214\,b, not corrected for telluric absorption, obtained with 
SofI on May 17/18, 2011.}
\label{spect}
\end{figure}

\begin{figure}[!t]
\centering{
\begin{tabular}{r}
\includegraphics[width=0.47\textwidth]{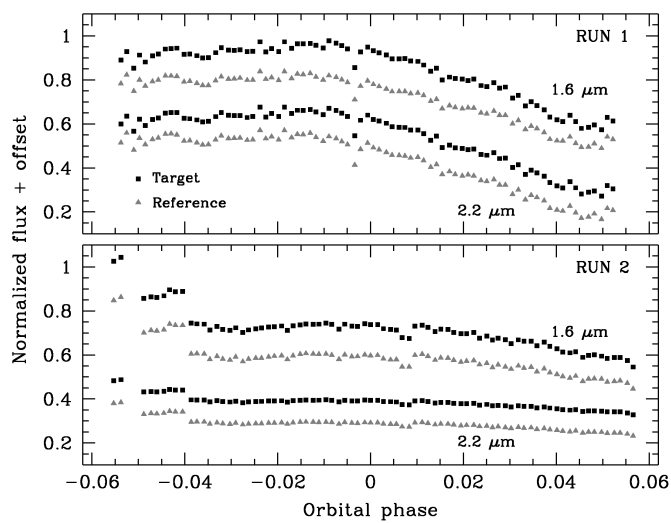}
\end{tabular}}
\caption{\label{fig:raw_spectra}Spectrophotometric individual time
  series obtained from integrating the flux of the {\it SofI} spectra
  within the $H$ and $K_S$ bands for the target and its comparison
  stars, and for the two good-weather runs. }
\end{figure}

\subsection{NIR spectrophotometry with SofI}

\begin{table*}[!htp]
\caption{Observing log for the spectroscopic SofI Observations of GJ\,1214\,b.}
\label{table:obs_log_SofI}
\begin{center}
\begin{tabular}{@{ }c@{ }c@{ }c@{ }c@{ }c@{ }c@{ }c@{ }c@{ }c@{ }c@{ }}
\hline
\hline
Date           &~Observations~& Predicted  &~Airmass range~& Cloud      &~Seeing~&~Slit Width~& Spectral   &~NDIT$\times$DIT~&~Number of~\\
~yyyy-mm-dd/dd~& UT range     &~Transit UT~& sec\,$z$      &~conditions~& arcsec & arcsec     &~Resolution~& N$\times$sec    & frames    \\
\hline
2011-05-17/18  & 5:44--9:55   & 7:32--8:26 & 1.2--2.1      & clear      & 0.8    & 1.0        & 600        & 3$\times$60     & 79        \\
2011-06-13/14  & 2:36--6:53   & 4:21--5:14 & 1.4--1.2--1.5 & clear      & 1.0    & 2.0        & 300        &~4$\times$45, 5$\times$36~& 79        \\
2011-08-09/10  & 0:45--4:07   & 1:49--2:42 & 1.2--2.0      & cirrus     & $>$1.0 & 2.0        & 300        & 3$\times$60     & 65        \\
\hline
\end{tabular}
\end{center}
\end{table*}

\begin{figure}[t]
\centering{
\begin{tabular}{r}
\includegraphics[width=0.40\textwidth]{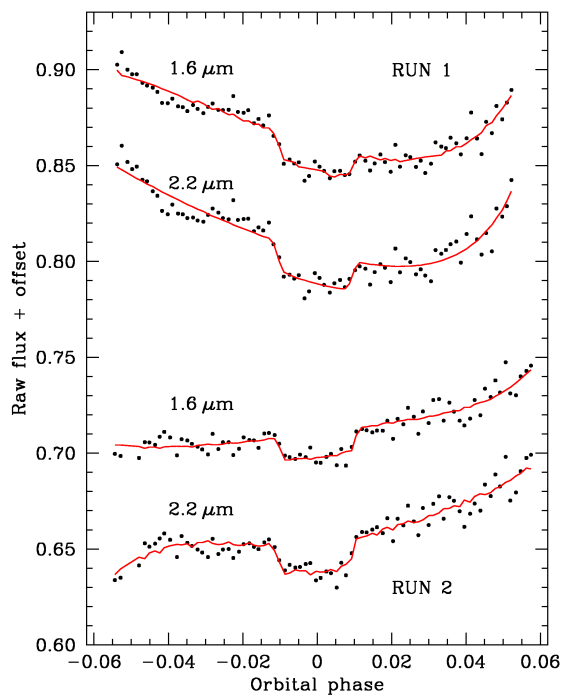}\\
\includegraphics[width=0.40\textwidth]{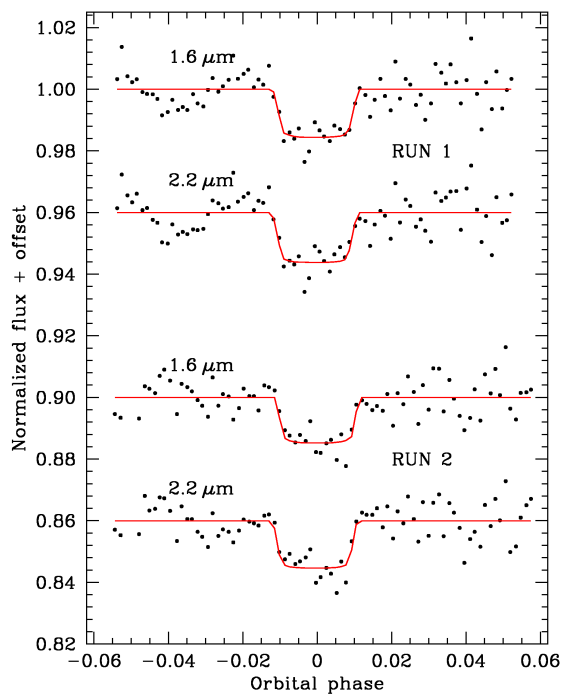}
\end{tabular}}
\caption{\label{fig:SofI_LCs} $H$ (1.6\,$\mu$m) and $K_S$
  (2.2\,$\mu$m) GJ\,1214\,b spectrophotometric series obtained from
  integrating the flux of the {\it SofI} spectra within the $H$ and
  $K_S$ bands. {\sl Top:} Raw light curves, with the best-fitting
  transit models in red. {\sl Bottom:} De-trended light curves with the
  best-fitting models. A color version of this plot can be found in
  the electronic version of the paper. }
\end{figure}

Three GJ\,1214\,b transits were observed with the NIR spectrograph
{\it SofI} \citep[Son of ISAAC;][]{moorwood_etal_1998} at the ESO New
Technology Telescope (see Table\,\ref{table:obs_log_SofI}). The
instrument is equipped with a 1024$\times$1024 Hawaii detector and
$\sim$4.9\,arcmin long slits (pixel scale 0.292\,arcsec\,px$^{-1}$).
The red grism was used, providing a spectral coverage from
$\lambda$$\sim$1.5 to 2.5\,$\mu$m.

A comparison star (2MASS J17152424+0455041; $\sim$3.06 arcmin
separation from the target) of similar spectral type and brightness as
GJ\,1214 was placed in the slit at all times for continuous and
simultaneous monitoring of the telluric absorption. The observations
were performed in stare mode, that is, without nodding, to keep the
objects on nearly the same pixels, minimizing effects from
flat-fielding errors and intra-pixel sensitivity variations. The
NDIT$\times$DIT combination was chosen to keep the peak count level at
$\sim$6000\,ADU, well below the $\sim$1.5\% nonlinearity limit of
10,000\,ADU.

The first steps of the basic data reduction were the usual cross-talk
and flat-fielding corrections, and bad-pixel replacement. However, the
stare mode made it impossible to remove the sky emission by
subtracting corresponding nodding image pairs, as is normally
done. Instead, we took advantage of having small field distortions and
bright targets to individually trace their continuum on each frame,
and to extract at each side of the objects two adjacent sky emission
spectra. Then, we subtracted from each object's one-dimensional
spectrum the linearly interpolated value of the two one-dimensional sky
spectra that correspond to the location of the object. Since the sky
spectra also include the dark and bias contributions, we successfully
removed the detector pattern from our data as well. These steps were
performed with the task {\it apall} from the IRAF package {\it
  twodspec}.

The wavelength calibration is based on xenon lamp spectra--for each
object and each frame we separately extracted a one-dimensional lamp
spectrum from the lamp frame using the same tracing and extraction
aperture as for that object on that frame. Typically, nine Xe lines were
used in the calibration, and the RMS was 1-2\,\AA. A typical target
spectrum at this stage of the processing is shown in Fig.~\ref{spect}.

The telluric absorption was removed by dividing in the wavelength
space the planet's host spectrum by the spectrum of the reference
star.  This was done separately for each individual frame, ensuring
that the absorption variations are accounted for, because the two
spectra were obtained simultaneously and very close on the sky.
 
Finally, we integrated the flux in the reduced spectra within selected
bandpasses -- the standard SofI NIR broad-band filters, $H$
($\lambda_C$=1.653\,$\mu$m, FWHM=0.297\,$\mu$m) and $K_S$
($\lambda_C$=2.162\,$\mu$m, FWHM=0.275\,$\mu$m). We refrained from
splitting the data into finer spectral bins because of the low
signal-to-noise ratio of the data. The individual spectrophotometric
series for GJ\,1214\,b and its comparison star are shown in
Fig.\,\ref{fig:raw_spectra}, and the composed light curves are shown
in Fig.\,\ref{fig:SofI_LCs} (top).

\section{Data analysis\label{sec:analysis}}
\begin{figure}[!t]
\begin{center}
\includegraphics[width=0.47\textwidth]{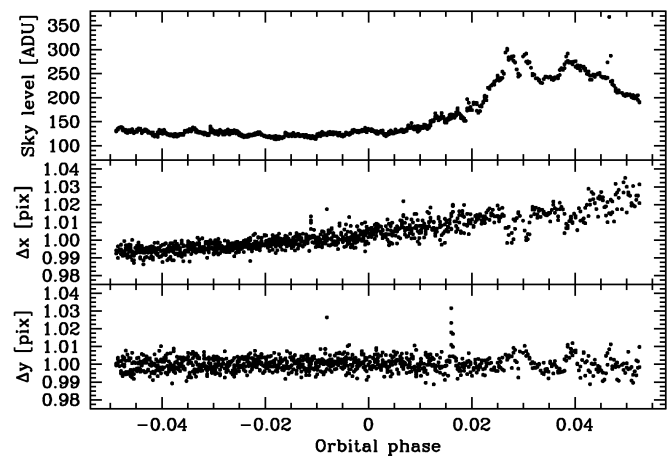}
\end{center}
\caption{\label{fig:center_sky} Parameters used to de-trend the OSIRIS
  photometric time series. {\sl From top to bottom:} Sky flux value
  and $(\Delta x_C,\Delta y_C)$, the variations of the target's
  position coordinates around their median value.}
\end{figure}

\subsection{{\it OSIRIS} and {\it SOI} data}
Despite the intrinsic correction of systematics due to differential
nature of the photometry, the {\it OSIRIS} raw light curve of
GJ\,1214\,b shows significant structure, especially in the second half
of the observations (Fig.\,\ref{fig:osirislc}, top). We found strong
correlations of the out-of-transit flux measurements with both the
$(x_C,y_C)$ position of the stellar centroids on the detector, and the
sky flux level $m_{s}$ -- related to guiding errors, residual
flat-field errors, and to variable weather conditions
(Fig.\,\ref{fig:center_sky}).

\begin{figure}[!t]
\begin{center}
\includegraphics[width=0.47\textwidth]{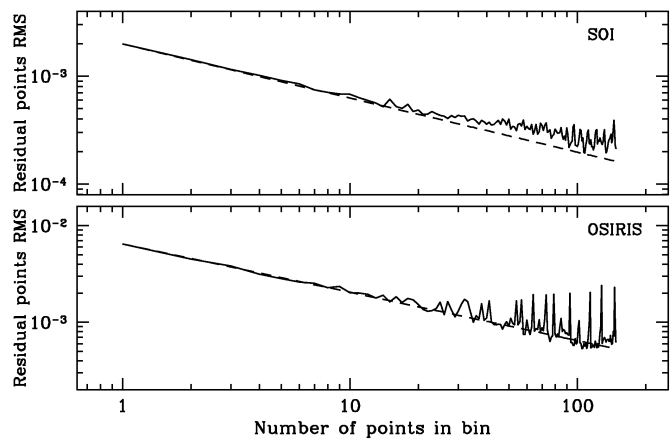}
\end{center}
\caption{\label{fig:rednoise} Red-noise diagram for the residuals of 
the best-fitting model for {\it SOI} (top) and {\it OSIRIS} (bottom). Solid 
lines represent the RMS calculated for a given bin width. Dashed 
lines show the expected behavior of the residuals for a pure 
Poisson-noise model.}
\end{figure}

To extract the physical parameters from the obtained light curves we
fit the photometric time series with both a transit model $\Theta
(t)$, and a linear model $\kappa (t)$ that accounts for the
systematics observed in the light curve. The best combination of
parameters we found to de-trend the {\it OSIRIS} light curve was
\begin{equation}
\kappa_{\textrm{\small OSIRIS}} (t) = a_0 + a_1\,m_s + a_2\,x_C + a_3\,y_C \,\, ,
\end{equation}
where the coefficients $a_i$ are the free parameters to be determined
in the fitting procedure.

The SOI data showed a prominent linear drift in the differential light
curve (Fig.\,\ref{fig:soilc}, top), which is most likely caused
by movement of the stars across the detector through the night. We
modeled it as\begin{equation}
\kappa_{\textrm{\small SOI}} (t) = a_0 + a_1\,x_C + a_2\,y_C \,\, .\end{equation}

The final fitted model for each data set was the product
\begin{equation}
M(t) = \kappa (t) \, \Theta(t) \,\, .
\end{equation}

To find the best-fitting model we carried out a Markov chain Monte
Carlo (MCMC) simulation \citep{tegmark_2004,Gregory_2005,Ford_2006},
aimed at drawing the a posteriori probability distribution for the
parameters to fit. It is widely used to determine physical parameters
from observations in exo-planetary science \citep[e.g.,][]{ford_2005,
  holman_etal2006,collier-cameron_etal2007}. The MCMC code uses a
Metropolis-Hasting algorithm to explore the parameter space.  The
GJ\,1214\,b planetary system model included a transiting planet
orbiting a star on a Keplerian orbit.

To create the transit model we made use of the quadratic
limb-darkening eclipse parametrization of \citet{mandel_agol2002}.
The free parameters in the fitting procedure,  jumps in the MCMC
code, are the time at the minimum flux $T_C$, the planet-to-star area
ratio $p^2$=$(R_p/R_s)^2$, the impact parameter $b$, the transit
length $T_{14}$, and the two quadratic limb-darkening coefficients
$u_1$ and $u_2$. However, we replaced the two limb-darkening
coefficients with their linear combinations
$c_1$=2$\times$$u_1$+$u_2$, and $c_2$=$u_1$-2$\times$$u_2$ because
this substitution has been proven to minimize the correlations among
the parameters in the MCMC fitting
\citep[e.g.][]{winn_etal2008,gillon_etal2010}

To determine the limb-darkening coefficients ($u_1$ and $u_2$), we
assumed Gaussian priors for the fitting procedure.  The starting
values were interpolated from the \citet{claret_bloemen_2011} tables
for the $K$ and $I$ band, and assuming the stellar parameters
$T_{\rm{eff}}$=3026.0$\pm$130.0\,K, $\log g$=4.991$\pm$0.029, and
[Fe/H]=0.39$\pm$0.15 \citep{charbonneau_etal2009}. We calculated the
errors in the limb-darkening parameters following
\citet{gillon_etal2009}. We adopted the orbital period
$P$=1.5803925\,d from \citet{charbonneau_etal2009} and fixed the
eccentricity to zero.

The linearity of the detrending models allowed us to calculate the
fitting coefficients with linear least-squares minimization using the
SDV algorithm \citep{press_1992} on the resulting curve produced
after dividing the raw light curve by the proposed transit model for
each specific jump \citep{gillon_etal2010}, instead of perturbing the
coefficients within the MCMC code.

We first ran five chains with $10^6$ jumps each, where we discarded the
first $20\%$ of points of each chain to remove the burn-in phase of
the simulation. Using the obtained best-fitting model of this first
run, we determined the significance of the red-noise in the photometric
time series by measuring the $\beta$ parameter defined by
\citet{gillon_etal2006} and \citet{winn_etal2008} as\begin{equation}
\beta = \frac{\sigma_N}{\sigma_1}\left(\frac{N(M-1)}{M}\right)^{1/2},
\end{equation}
where $M$ is the obtained number of bins for a specific bin width $N$,
and the $\sigma_N$ and $\sigma_1$ parameters represent the RMS of the
binned and unbinned residuals obtained after removing the best-fitting
model, respectively. We selected the value that corresponds to a
temporal scale similar to the ingress-egress duration, that is,
$\sim$6\,min \citep{carter_etal2011}, which yielded values of
$\beta_\textrm{\small OSIRIS}$=1.03 and $\beta_\textrm{\small
  SOI}$=1.23. We multiplied the individual photon-noise flux errors by
this value to execute a new set of chains including the red-noise.

With the new set of flux errors, we ran five new MCMC chains with the
same configuration as described above to obtain the final a posterior
probability distribution of the jump parameters. The final values for
each of the jump parameters were the median of the distribution, and
its errors correspond to the boundaries of the region enclosing the
$68.3\%$ of values around the median. A \citet{gelman_rubin1992} test
was applied to this MCMC run to check for a good mixing and
convergence, which yielded in a successful result. The final simulated
and derived parameter values are presented in Table
\ref{table:mcmc_parameters}.

\begin{table}[!t]
\caption{Transit parameters of GJ\,1214\,b derived from the {\it OSIRIS} 
and SOI data.}
\label{table:mcmc_parameters}
\begin{tabular}{cccc}
\hline
\hline
Parameter & Value & 68.3\% Conf. Limits & Unit\\
\hline
\multicolumn{4}{c}{}\\
\multicolumn{4}{c}{\sl OSIRIS - 2.14\,$\mu$m}\\
$T_C$          & 2455783.59415 & $-0.00025$, $+0.00024$ & BJD  \\
$R_p/R_s$      & 0.1173    & $-0.0024$, $+0.0022$ &  \\
$u_1$     & 0.032     & $-0.011$, $+0.008$   &\\
$u_2$     & 0.230     & $-0.010$, $+0.011$   &\\
\multicolumn{4}{c}{}\\
\multicolumn{4}{c}{\sl SOI - $I$-Bessel - 0.8665\,$\mu$m}\\
$T_C$          & 2455315.79475 & $-0.00010$, $+0.00011$ & BJD  \\
$R_p/R_s$      & 0.11735   & $-0.00076$, $+0.00072$   & \\
$u_1$     & 0.341     & $-$0.010$, $+0.011 &\\
$u_2$     & 0.301     & $-$0.011$, $+0.010 &\\
\multicolumn{4}{c}{}\\
\multicolumn{4}{c}{\sl Common parameters}\\
$P$        & 1.580404938 & $\pm 0.000000090$  & days \\
$T_0$      & 2454934.916934 & $\pm 0.000027$ &  BJD\\
$a\,^*$    & 14.9749     & -   & $R_S$ \\
$i\,^*$    & 88.94    & - & deg \\
$e\,^*$    & 0.0       & -                & \\
$\omega\,^*$    & 0.0       & -                & deg \\
\hline
\end{tabular}\\
\begin{minipage}[t]{0.9\columnwidth}
\medskip
$^*$ Value fixed in the MCMC fitting~\citep{bean_etal2010}.
\end{minipage}
\end{table}

The determined parameters agree well with those reported by
\citet{bean_etal2011} and \citet{carter_etal2011}. We decided to
decrease the number of free parameters in our MCMC fitting by fixing
the values given by \citet{bean_etal2011}, which are $i = 88.94^\circ$
and $a/R_{S} = 14.9749$. This selection allowed us to directly compare
the measured depth values with those reported in
\citet{desert_etal2011}, \citet{croll_etal2011},
\citet{bean_etal2011}, \citet{demooij_etal2012},
\citet{murgas_etal2012}, \citet{narita_etal2013,narita_etal2012},
\citet{fraine_etal2013}, and \citet{teske_etal2013}.

The final de-trended light curves, the best-fitting transit-plus-trend
models, and their residuals are shown in the middle and bottom panels
in Figs.\,\ref{fig:osirislc} and \ref{fig:soilc} for the OSIRIS and
SOI observations, respectively. Low time-correlated noise in the SOI data is apparent in
Fig.\,\ref{fig:rednoise} (top).

\begin{figure}[!t]
\begin{center}
\includegraphics[width=0.24\textwidth]{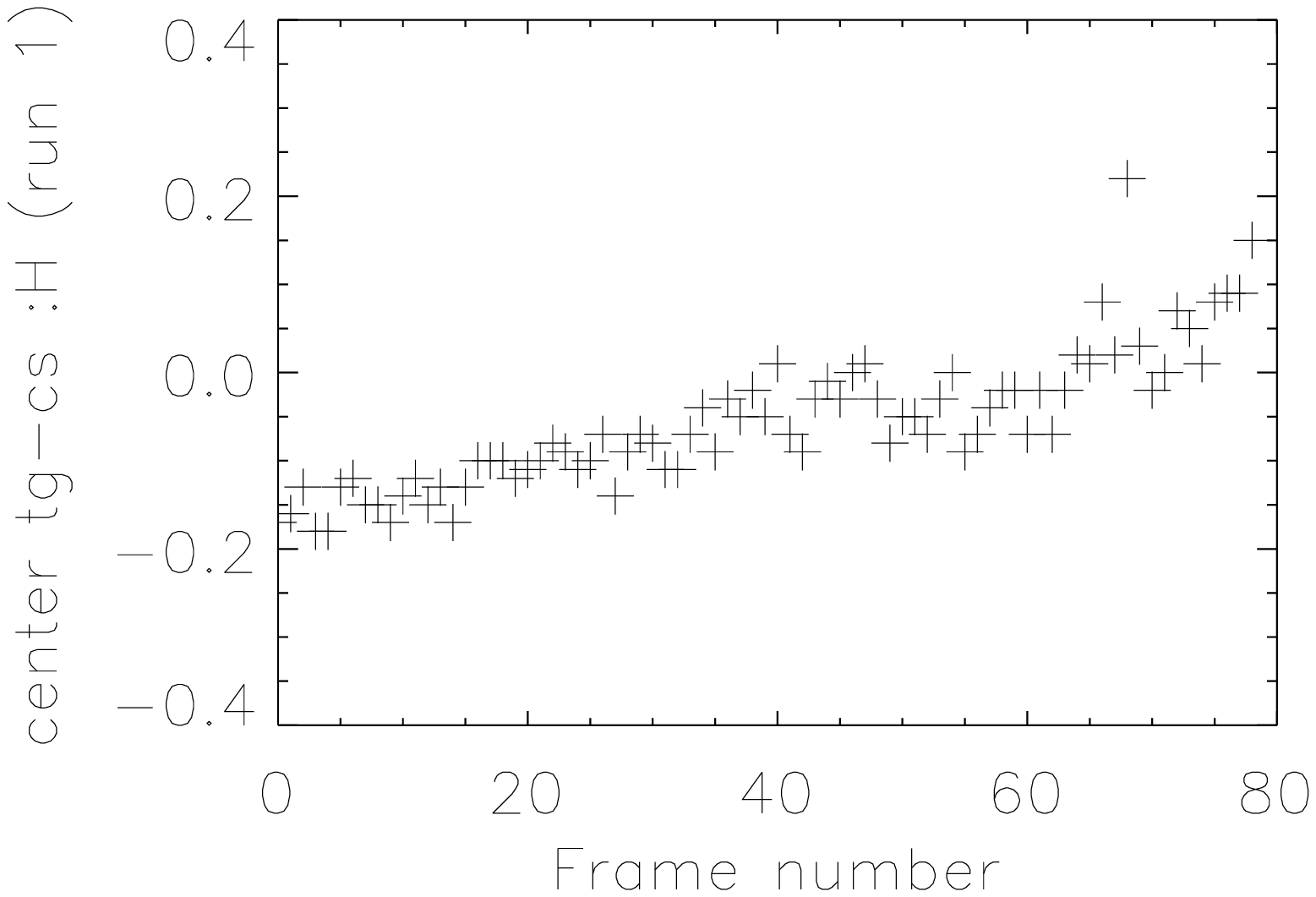}
\includegraphics[width=0.24\textwidth]{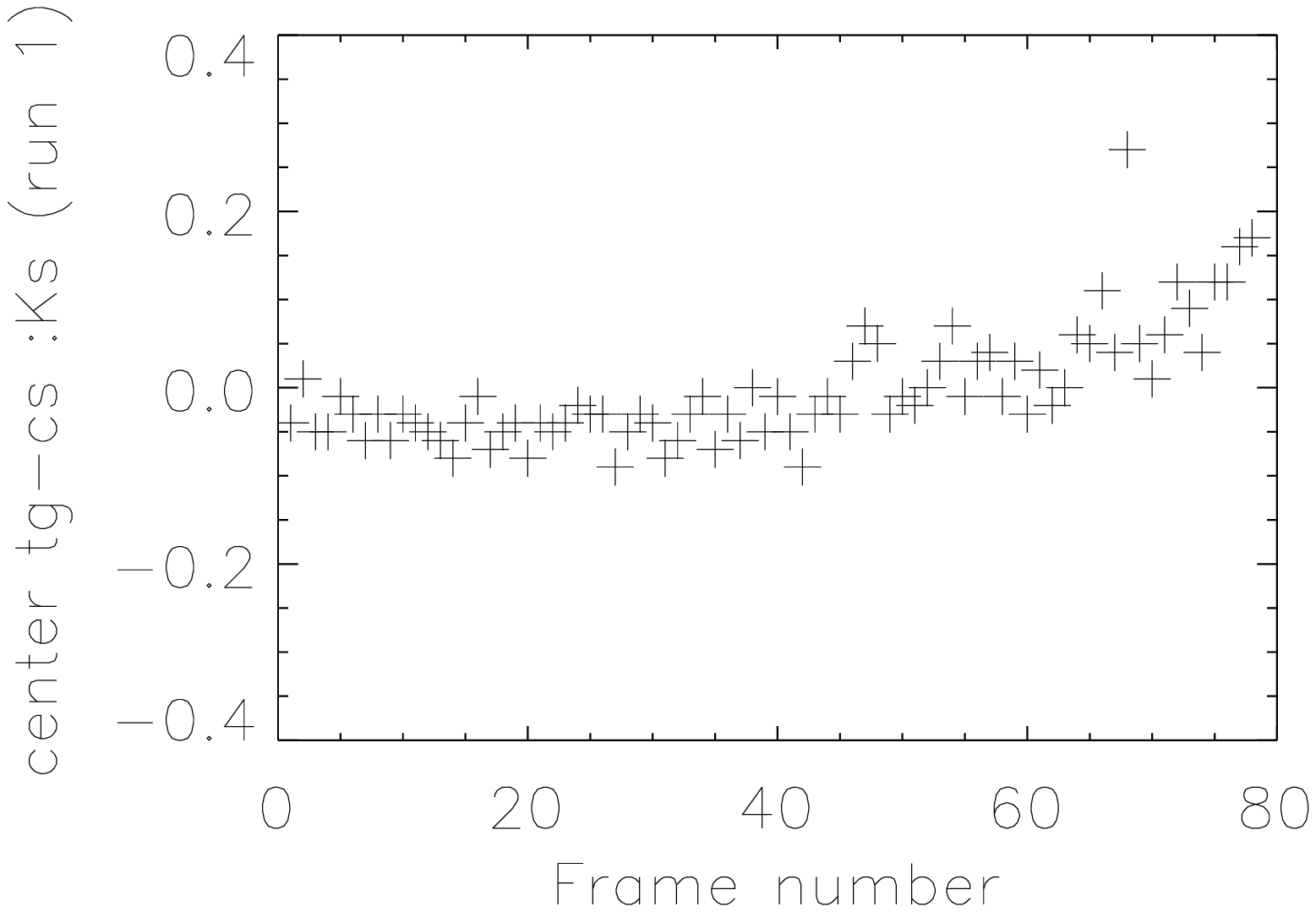}
\includegraphics[width=0.24\textwidth]{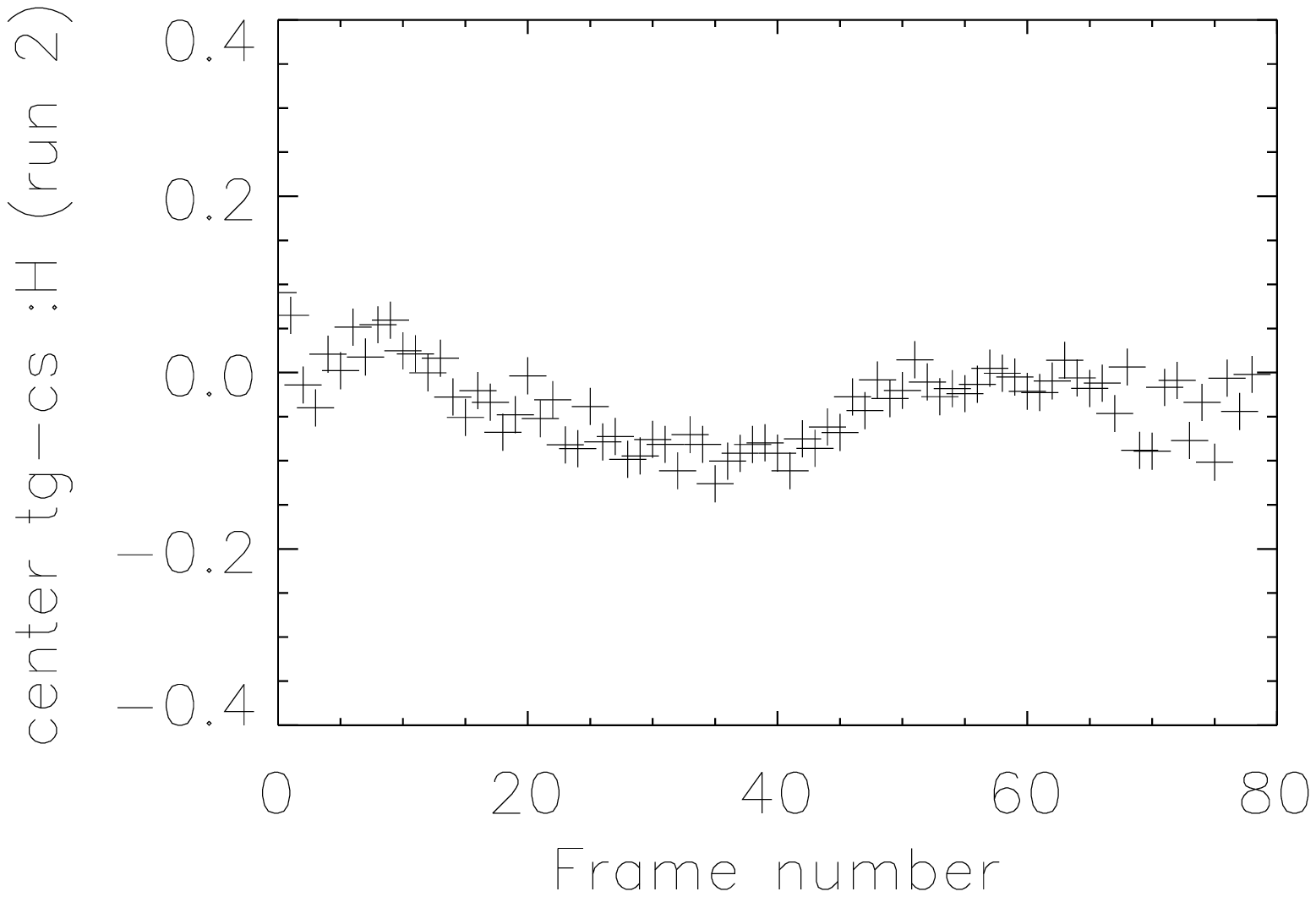}
\includegraphics[width=0.24\textwidth]{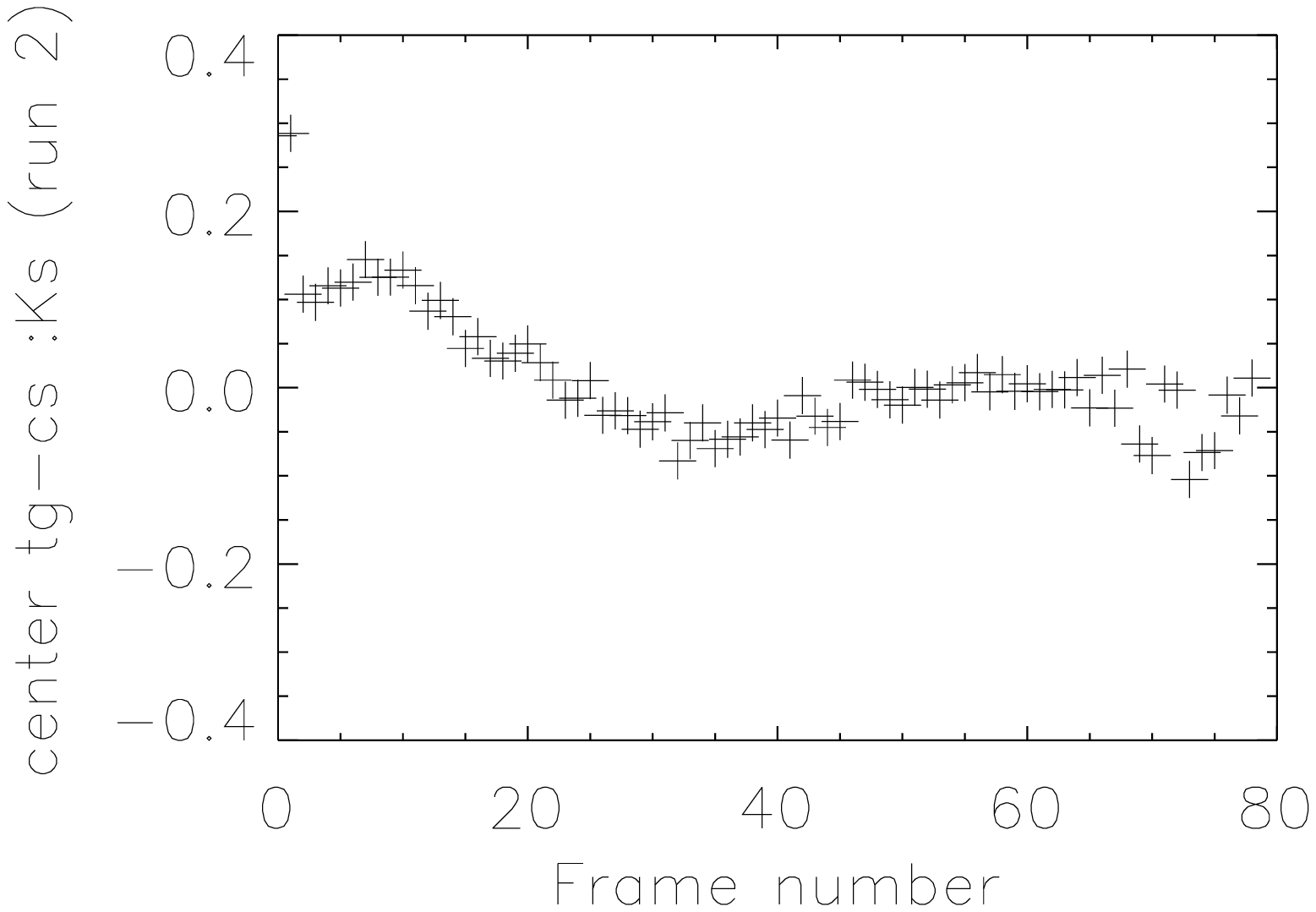}
\includegraphics[width=0.24\textwidth]{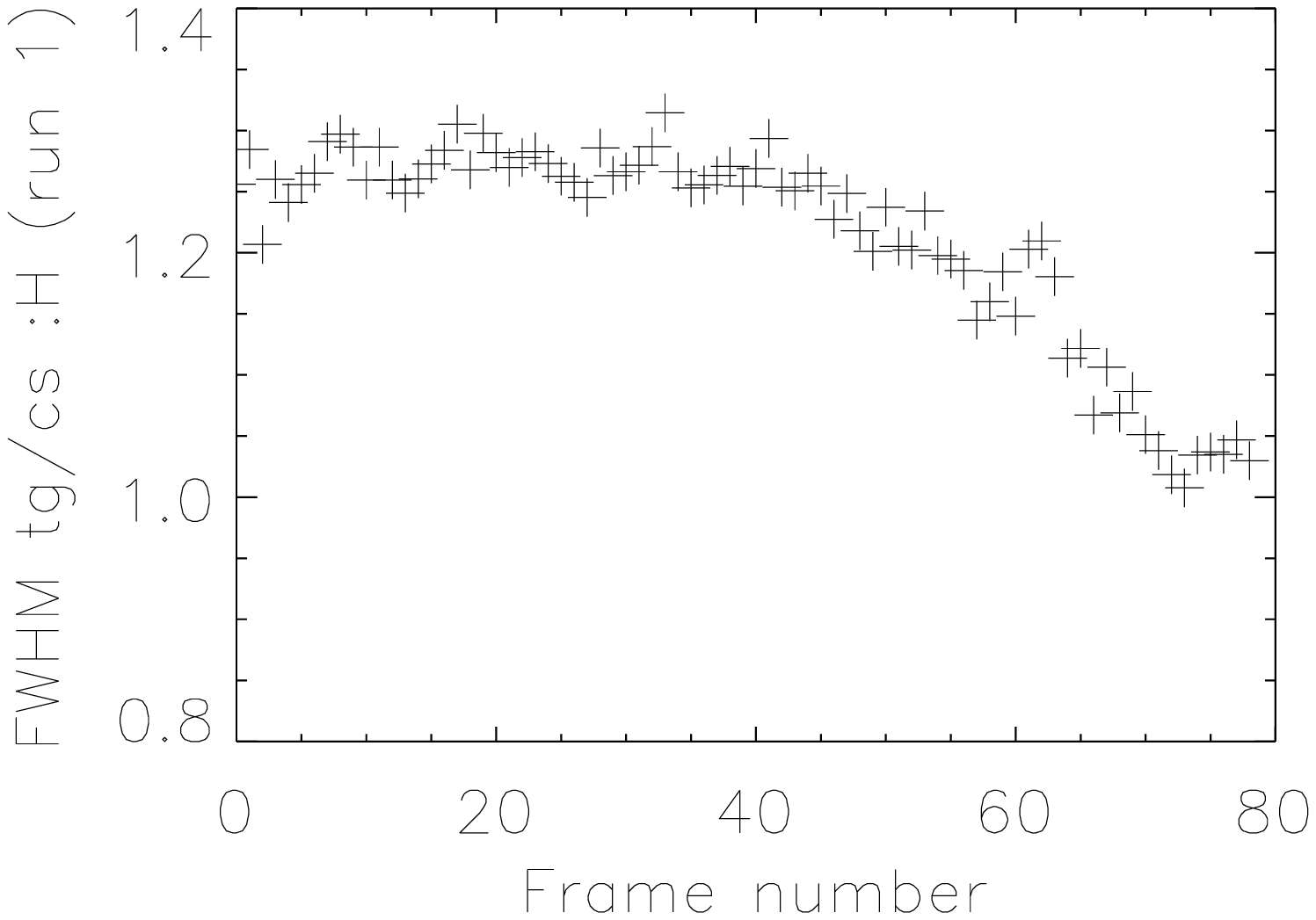}
\includegraphics[width=0.24\textwidth]{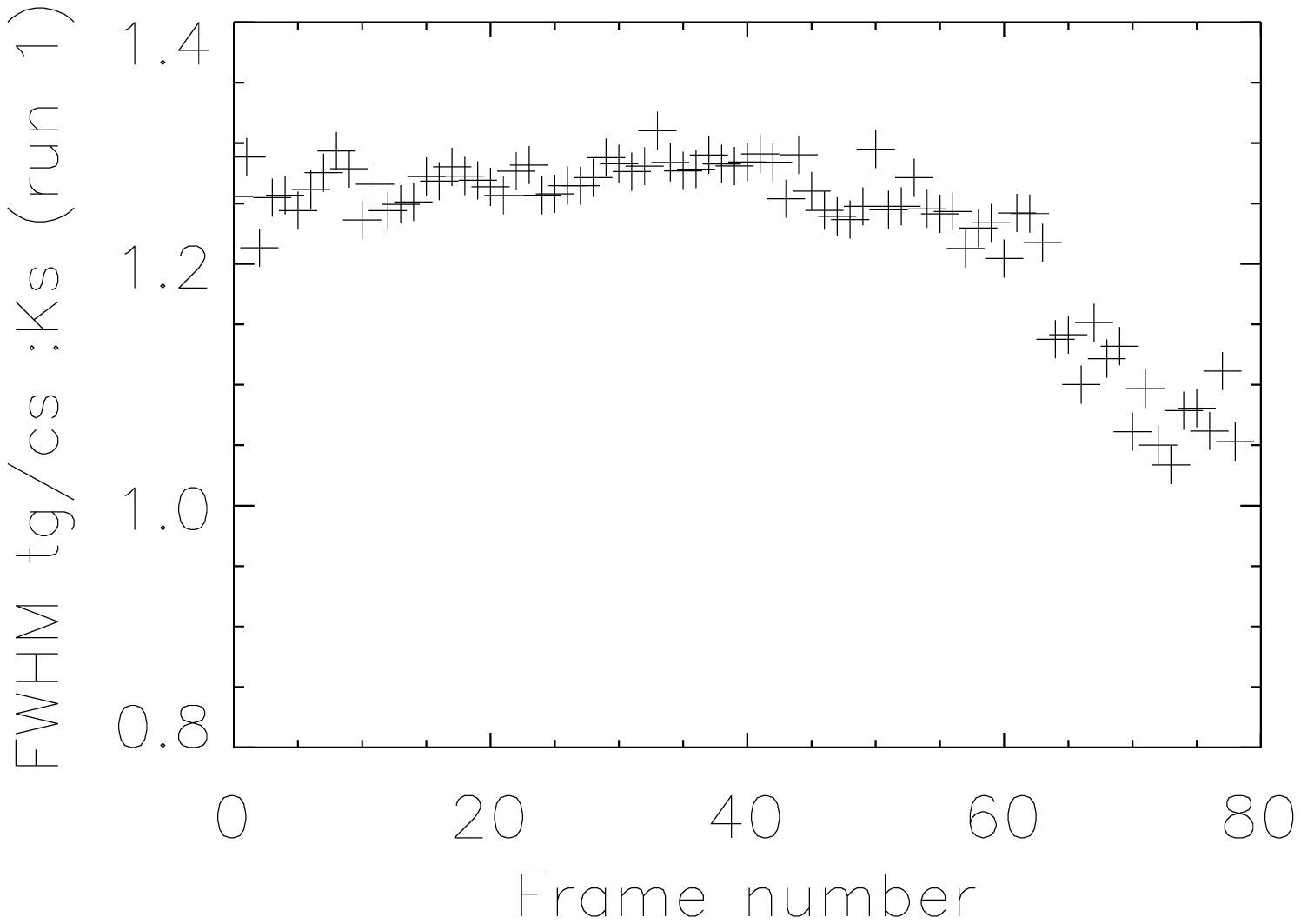}
\includegraphics[width=0.24\textwidth]{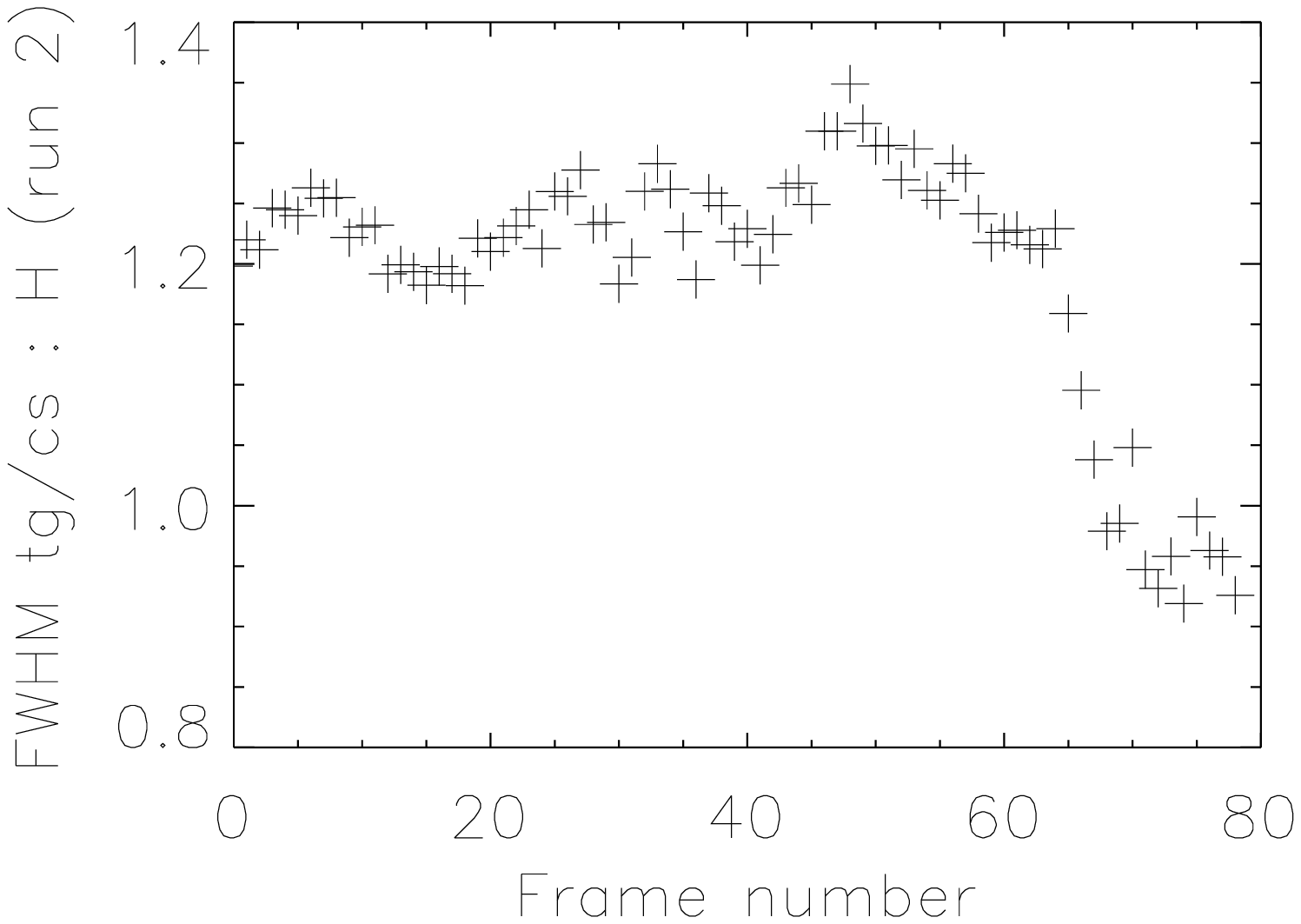}
\includegraphics[width=0.24\textwidth]{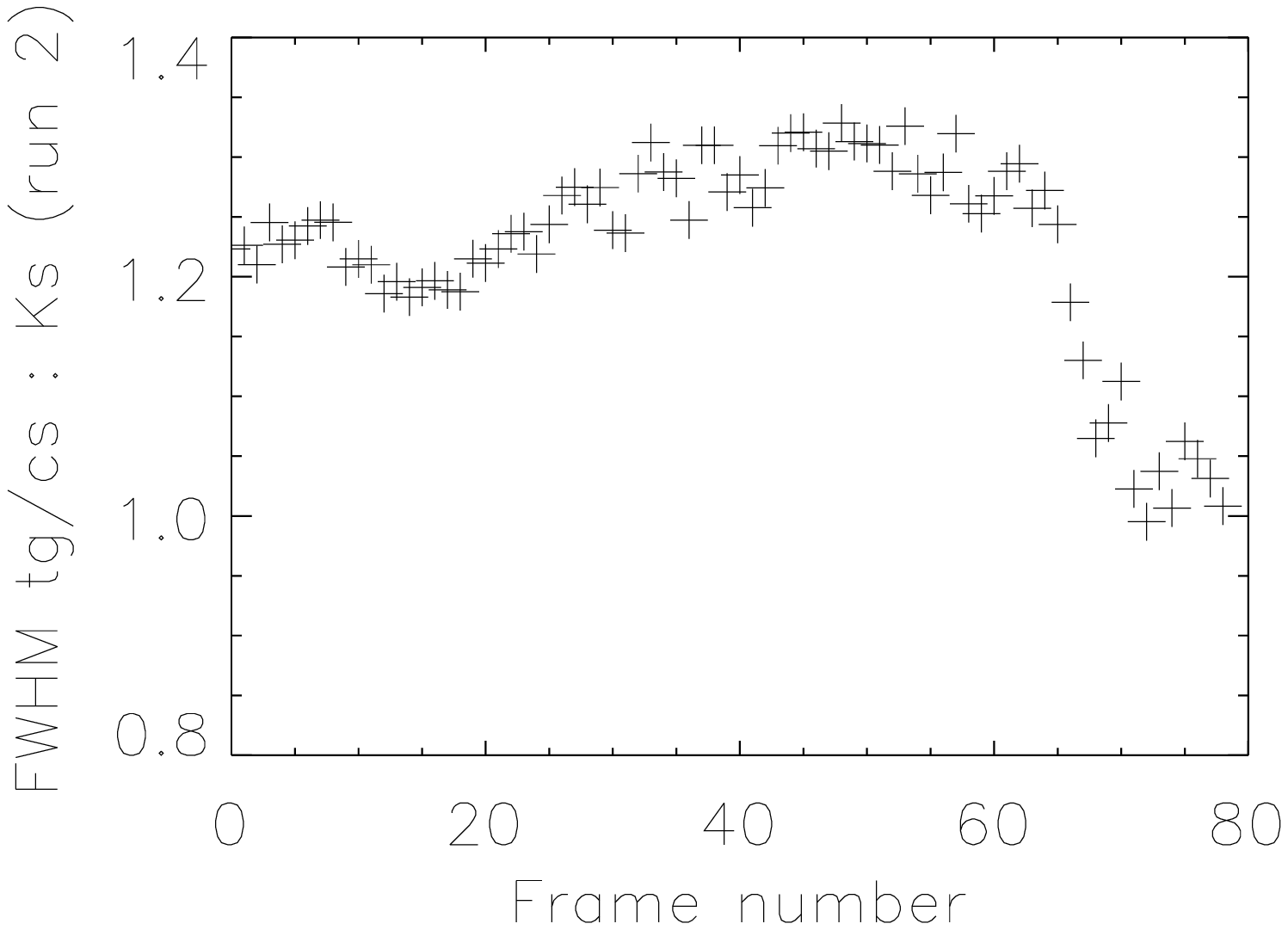}
\end{center}
\caption{\label{fig:FWHM}First set of graphs: changes
  in the distance between the target and comparison along the slit
  for two runs and two different wavelengths corresponding to center
  of the $H$ and $K_S$ filters. The second set of plots depicts a ratio of
  FWHMs of the target and the comparison for first two runs for center
  pixels of $H$ and $K_S$ wavelengths.}
\end{figure}

\subsection{Analysis of {\it SofI} data}

First, we studied the main sources of systematic errors.  One of our
concerns was to understand how stable the centering of the target and
the reference sources on the slit is. We measured the drifts and the
FWHM variations of both objects at two different wavelengths,
corresponding to the $H$ and $K$ bands.  The analysis consisted of the
following steps: (i) determine the pixel corresponding to the center
of the band, (ii) fit the spectral profiles at these wavelengths with
Gaussian profiles to determine the center and FWHM of the two spectra.
These steps were repeated independently for each good-weather run. The
results for these runs are plotted in Fig.~\ref{fig:FWHM}.

\begin{figure}[!t]
\begin{center}
\includegraphics[width=0.48\textwidth]{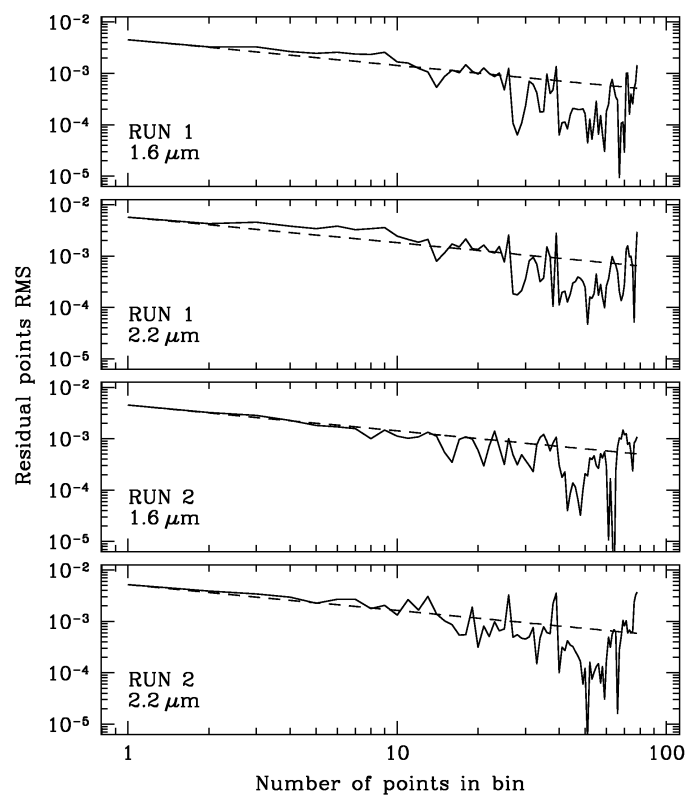}
\end{center}
\caption{\label{fig:rednoisesofi}Red-noise diagram for the residuals of 
the best-fitting model for SoFI data for two runs and two different wavelength channels. Solid 
lines represent the RMS calculated for a given bin width. Dashed 
lines show the expected behavior of the residuals for a pure 
Poisson-noise model.}
\end{figure}

Apparently, the target-to-reference separation on the detector along
the slit is stable to within 0.25 pixels ($\sim$0.007”). We were
unable to measure the positions of the stars across the slit with the
same accuracy. Nevertheless, if we assume that the objects show the
same relative drift in that direction as well, then the change of the
flux ratio for two stars of similar brightness in the most extreme
case, corresponding to a drift of 0.007” for one star while the other
stays well centered on the slit, is about 0.2\% for the whole run. The
first run is slightly more stable. However, the monitoring of the
separation along the slit indicates that it varies smoothly along the
slit, without jumps, particularly, without dramatic jumps at the time
of ingress and egress. This smooth variation is likely to occur due to
color-dependent differential atmospheric refraction or smooth guider
drift -- mechanisms that are probably not related with the slit
orientation. Therefore, we expect that the drift across the slit
follows a similar smooth pattern.

The FWHM ratio (Fig.\,\ref{fig:FWHM}, bottom) also varies smoothly
with time and is distinctly different from unity. The latter is
related to the fixed position of the SofI collimator - for technical
reasons it was not moved to adjust the focus for each set up. Instead,
it was set at a position optimized for imaging observations, and the
focus at spectroscopy is suboptimal, leading to a gradient across the
field of view. This is the reason why the FWHM is different at the
locations of the target and the reference stars. The effect of the
suboptimal focus is more pronounced at the beginning of the
observations at lower airmass, and weaker towards the end, when the
higher airmass smears the images and makes the de-focusing less
important.

The resulting light curve was analyzed via two different MCMC-based
routines: one as described in previous section for our photometric
data sets, and another based on the code by \citet{gillon_etal2009}
and \citet{gillon_etal2012}.  Both methods provided consistent
results. For clarity we present in the following text and in
Table\,\ref{res:sofi} the fitting parameters obtained with the second
mentioned MCMC code described in \citet{gillon_etal2009} and
\citet{gillon_etal2012}. The RMS of the residuals for different bin sizes
are presented in Fig.~\ref{fig:rednoisesofi}. The transit is very short
and the ingress/egress phase lasts only a few minutes. Therefore, the
comparable bin size with comparable length of ingress/egress
corresponds to binning of 2 or 3 points. Figure~\ref{fig:rednoisesofi}
suggests that the red-noise contribution to the overall error budget
is fairly low. However, due to time sampling, it is not possible
to draw a more accurate conclusion.

\begin{figure*}[!t]
  \begin{center}
     \includegraphics[width=0.95\textwidth]{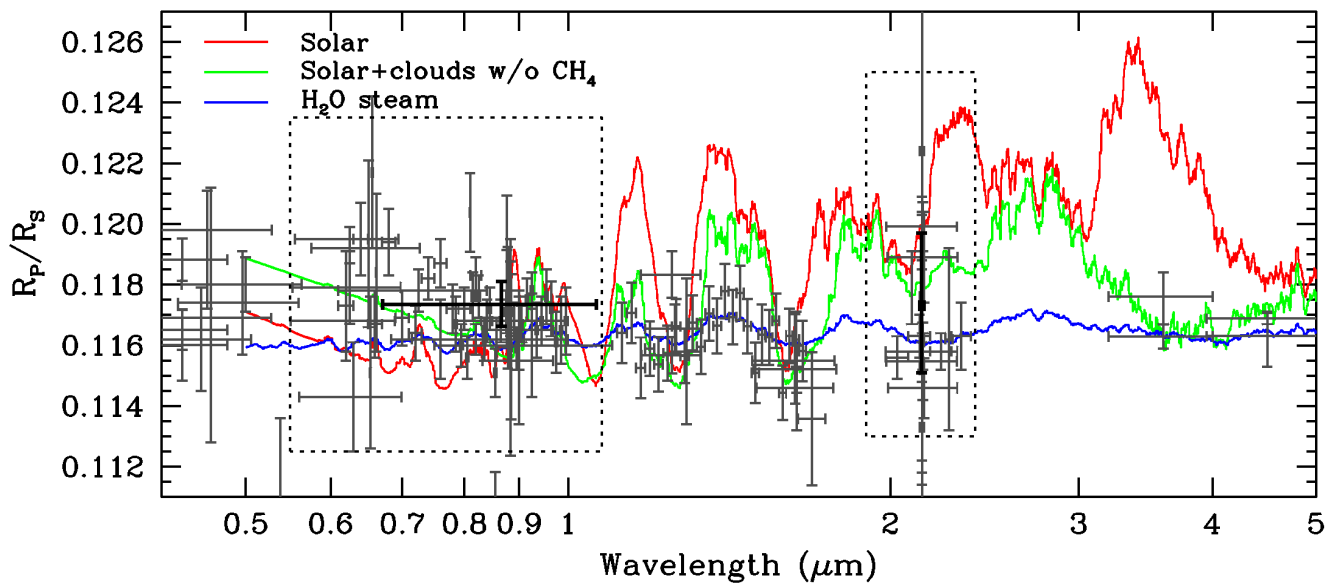}
  \end{center}
  \caption{\label{fig:models} Updated best fit atmospheric models
    reported in \citet{miller-ricci_etal2012}, including all current
    observational data. The models shown were smoothed for the sake of
    clarity. Light-gray points represent spectrophotometric
    measurements, and dark-gray points represent photometric
    measurements. The horizontal extension of the bars represents the
    wavelength coverage of the point, instead of an error associated
    with the observation. Black points represent our {\it SOAR}
    measurements. The dashed windows represent the zoomed-in regions
    displayed in Fig.~\ref{fig:zoomin}. A color version of this plot
    can be found in the electronic version of the paper.}
\end{figure*}

\begin{table}[ht]
\caption{Transit parameters of GJ\,1214\,b obtained from the {\it SofI} data}
\label{res:sofi}
\begin{tabular}{cccc}
\hline
\hline
Parameter & Value & 68.3\% Conf. Limits & Unit\\
\hline
\multicolumn{4}{c}{}\\
\multicolumn{4}{c}{\sl SofI H+K run 1}\\
$T_C$        &    2455699.83386 & $-0.00093$, $+0.00094$  & BJD\\
$R_p/R_s$    &    0.1205  &  $-0.0074, +0.0070$  &  1.6\,$\mu$m\\
$R_p/R_s$    &    0.1230  &  $-0.0090, +0.0084$  &  2.2\,$\mu$m\\

\multicolumn{4}{c}{}\\
\multicolumn{4}{c}{\sl SofI H+K run 2}\\
$T_C$        &    2455726.7001 &  $-0.0010$, $+0.0010$ &  BJD \\
$R_p/R_s$    &   0.1180  &  $-0.0062$, $+0.0058$  & 1.6\,$\mu$m\\
$R_p/R_s$    &   0.1202  &  $-0.0110$, $+0.0100$  & 2.2\,$\mu$m \\
\hline
\end{tabular}
\end{table}

\section{Discussion\label{sec:disc}}
\subsection{Spectrophotometric observations}

The careful analysis shows that SOFI is fairly stable in terms of
differential flux losses, and the duty cycle was extremely high --
95-97\%, that is, we lost only 3-5\% of the time for detector readout,
fits-file merging, and transferring.  Unfortunately, the loss of the
third night because of poor weather conditions undermined the analysis
of our spectroscopic data, and the brightness of the exoplanet host
was marginal. We still present the data here as a benchmark and
reference for similar future observations, to demonstrate the
potential of long-slit spectrographs at 4-m class telescopes to study
extrasolar planets \citep[e.g.][]{crossfield_etal2013}.

SOFI has a major advantage over other spectrographs: its long slit
length of $\sim 5\arcmin$, which facilitates finding a suitable
comparison star, which is necessary to control the systematics (see
discussion in previous section and Fig.~\ref{fig:rednoisesofi}).
Therefore, we conclude that SOFI is suitable for studying exoplanet
atmospheres, if the planets orbit brighter stars.

\subsection{Photometric GJ\,1214\,b atmosphere}
A variety of measurements of the effective planet-to-star radius ratio
at various wavelengths have provided conflicting clues on the
composition of the GJ\,1214b's atmosphere. The first detections of
this planet suggested that its radius is too large to be explained by a
solid (pure rock or pure ice) composition
\citep[e.g.,][]{charbonneau_etal2009}, implying the presence of a
significant gaseous atmosphere. Depending on the assumptions made for
the composition of the planet's interior, this atmosphere could be
composed primarily of hydrogen, water, or some combination thereof,
a hypothesis that cannot be probed from a single radius measurement.

More recent detections at different wavelengths have suggested a high
molecular weight atmosphere with a probable dominance of water, which
would show shallow $K$-band depths \citep[e.g.,][]{bean_etal2010,
  desert_etal2011, bean_etal2011, berta_etal_2012}. At the same time,
\citet{croll_etal2011} reported a deep $K_S$-band transit, which
favors a low molecular weight atmosphere, with an hydrogen-rich
component; this result was accompanied by the relatively deep
$K_S$-band transit reported by \citet{demooij_etal2012}. Meanwhile,
\citet{crossfield_etal_2011} reported that the hydrogen-rich atmosphere is
not favored by their results.  As noted by \citet{howe_burrows2012},
the shorter wavelength measurements reported by \citet{bean_etal2011}
and \citet{demooij_etal2012} favor hydrogen-rich atmospheres as well.

\citet{nettelmann_etal2011} considered different models for
GJ\,1214b's interior, inferring metal-rich H/He atmospheres are the
most plausible models, and suggesting an H/He/H$_2$O model with a high
water mass fraction for the atmosphere of the planet.

\begin{figure*}[!t]
  \centering{ \includegraphics[width=0.47\textwidth]{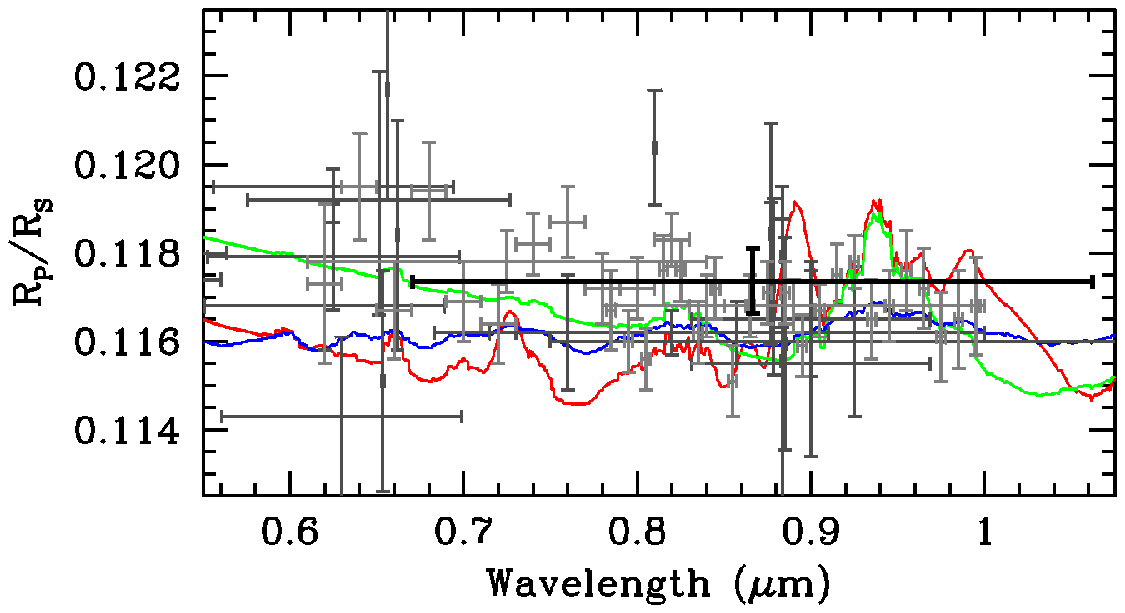}
    \hspace{20pt} \includegraphics[width=0.47\textwidth]{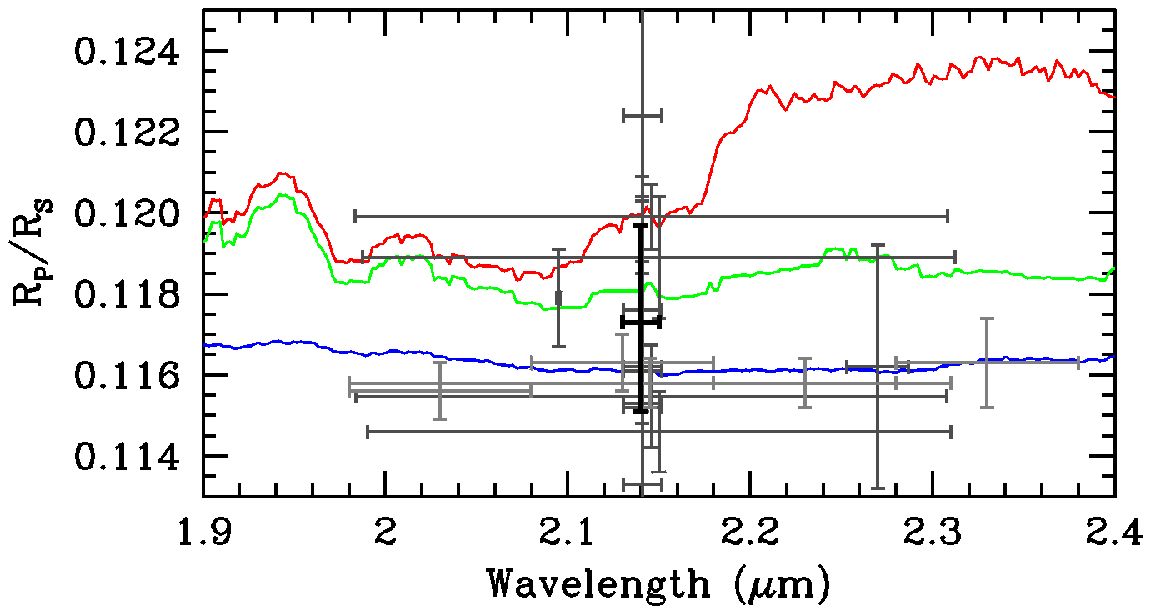}}
  \caption{\label{fig:zoomin} {\sl Left:} A zoom-in from
    Fig.~\ref{fig:models} for the optical region around our $I$-Bessel
    measurements. {\sl Right}: The $K$-band region of spectra around
    our $2.14\,\mu$m observation. Our measurement points are
    represented by dark circles, while gray points follow the
    description in Fig.~\ref{fig:models}. A color version of this plot
    can be found in the electronic version of the paper.}
\end{figure*}

There is no individual theoretical model that accounts for all the
observational data, but atmospheres that either have a water-rich
composition (more than 70\% by mass) or a thick layer of clouds or
hazes in the upper atmosphere are the best-suited interpretations of
current data \citep[e.g.,][]{fortney_2005, miller-ricci_fortney2010,
  nettelmann_etal2011, howe_burrows2012, miller-ricci_etal2012}. The
last option has been studied by \citet{morley_etal2013}, who found
that in an enhanced metallicity atmosphere, clouds that formed either
in chemical equilibrium or nonequilibrium frameworks can reproduce
current observations. The authors also pointed out that hydrocarbon
haze produced by photochemistry can flatten the GJ\,1214\,b spectrum.

Some studies have been performed to try to solve the
discrepancies between the models. \citet{murgas_etal2012} used GTC
tunable filters to attempt the detection of the H$\alpha$ signature
during transits of GJ\,1214b, which yielded a nondetection, consistent
with the featureless transmission spectra presented by
\citet{bean_etal2011}. Finally, \citet{kreidberg_etal2014} have provided strong
  evidence of the presence of clouds in the atmosphere of GJ\,1214\,b,
  based on HST transmission spectra. They reported the significant
  detection of a featureless spectrum, ruling out the cloud-free high
  molecular weight atmosphere hypothesis.

Figure~\ref{fig:models} shows all current observational data that have
a wavelength-dependent radius ratio, where the models correspond to
updated best-fit atmosphere models proposed in
\citet{miller-ricci_etal2012}. Photometric data were obtained from
\citet{demooij_etal2012}, \citet{carter_etal2011},
\citet{bean_etal2011}, \citet{murgas_etal2012},
\citet{croll_etal2011}, \citet{desert_etal2011},
\citet{narita_etal2013,narita_etal2012}, \citet{teske_etal2013},
  \citet{colon_gaidos2013}, \citet{wilson_etal2013},
  \citet{gillon_etal2013}, and \citet{demooij_etal2013}. Transmission
spectroscopy measurements were obtained from \citet{berta_etal_2012},
\citet{bean_etal2011}, and \citet{kreidberg_etal2014}. Our
measurement at $0.87 \mu m$ agrees well with current
measurements at the shorter wavelengths, also suggesting the presence
of the Rayleigh scattering tail argued in \citet{howe_burrows2012}. On
the other hand, our NIR detection at $2.14\,\mu m$ shows a moderate
depth that disagrees with that of \citet{croll_etal2011} and
the $K_S$-band detection of \citet{desert_etal2011}. Recently,
\citet{narita_etal2012} reported simultaneous $J$, $H$, and $K_S$-band
transit depths for transits of GJ\,1214\,b. Of particular interest is
their shallow detection at $2.16\,\mu m$, which strongly disagrees
with the deeper measurements. Our $2.14\,\mu m$ detection is
consistent with the detections of \citet{narita_etal2012} and
\citet{bean_etal2011} and the $K_C$ detection of
\citet{desert_etal2011}.%

For better orientation, Fig.~\ref{fig:zoomin}, left, is a zoom-in of
Fig.~\ref{fig:models} with a focus on the optical region around our {\it
  SOI} measurement, while the same Fig.~\ref{fig:zoomin}, right,
shows the near-IR region around our {\it OSIRIS} measurement. For both
panels, relevant measurements by the above mentioned groups are
presented for comparison.

Finally, we would like to point out a discrepancy between the
narrow-band and broad-band photometric measurements at similar
wavelengths (see Fig.~\ref{fig:zoomin}). Our measurement at
$2.14\,\mu$m contrasts with the broad band measurements
by~\citet{croll_etal2011} and \citet{narita_etal2012} and agrees well
with the narrow-band measurement of \citet{demooij_etal2012} at
$2.27\,\mu$m.

\begin{figure}[!pt]
  \begin{center}
     \includegraphics[width=0.96\columnwidth]{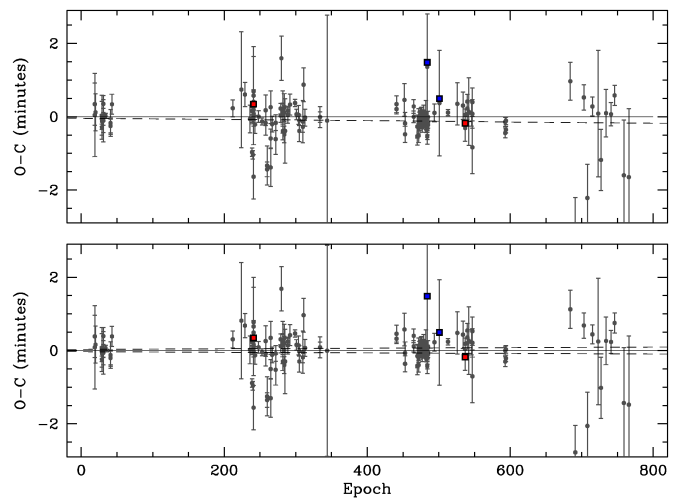}
  \end{center}
  \caption{\label{fig:timing} {\sl Top:} The observed-minus-calculated
    (O-C) diagram for the whole set of transit timing available to
    date in the literature, calculated on the ephemeris given by
    \citet{bean_etal2011}. Our measurements are drawn as filled
    squares and in addition highlighted by colors in the online
    version. The dashed line represents the new ephemeris calculated
    in this work. {\sl Bottom:} The same data set after removing the
    new ephemeris, including its 1-$\sigma$ errors (dashed lines).}
\end{figure}

\subsection{Transit-timing observations}
The timing information in the raw images was converted from MJD to BJD
(TDB) to determine the final individual transit timing, following the
prescriptions in \citet{eastman_etal2010}. Our photometric data show
no significant deviations from a constant period, which is consistent
with what has been found by \citet{carter_etal2011} and
\citet{berta_etal_2012}, and has been recently confirmed by
\citet{harpsoe_etal2013}, who used a Bayesian approach to infer that a
TTV is unlikely to be present in the GJ\,1214b data. We collected all
available transit-timing data from the literature, which we combined
with our measurements to refine the ephemeris of the GJ1214b system,
with parameters $T_0 = 2454934.917003 \pm 0.000023$\,BJD and $P =
1.580404599 \pm 0.000000056$. Timing data from
\citet{demooij_etal2012}, \citet{Kundurthy_etal2010},
\citet{carter_etal2011}, \citet{murgas_etal2012},
\citet{bean_etal2011}, \citet{charbonneau_etal2009},
\citet{berta_etal2011}, \citet{berta_etal_2012},
\citet{croll_etal2011}, \citet{desert_etal2011},
\citet{sada_etal2010}, \citet{narita_etal2012},
\citet{harpsoe_etal2013}, \citet{teske_etal2013},
  \citet{gillon_etal2013}, \citet{colon_gaidos2013},
  \citet{fraine_etal2013}, and our measurements are shown in
Fig.~\ref{fig:timing}.

\begin{figure}[t]
  \begin{center}
     \includegraphics[width=0.96\columnwidth]{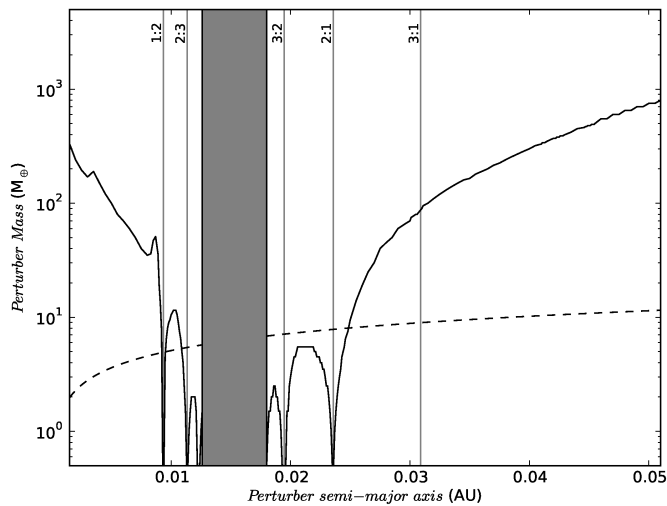}
  \end{center}
  \caption{\label{fig:MvsA} Upper mass limits for the system
    GJ\,1214b based on numerical simulations. The dashed line
    represents the limits imposed by the radial velocity
    measurements.}
\end{figure}

Based on the timing analysis of GJ1214b transits, we put strong
constraints on the mass of an additional body in the system,
especially in mean motion resonances (MMRs) with the transiting
exoplanet.  Using dynamical simulations with the MERCURY N-body
orbital integrator \citep{chambers_1999}, we determined the mass of an
orbital perturber as a function of the distance from the star. To run
the simulations we followed the same procedure as described in detail
in \citet{hoyer_etal2011}. We used the updated physical parameters of
the system reported by \citet{anglada-escude_etal2012} as input for
the simulations. For the perturber body, we explored a wide range of
masses ($0.5$\,$M_{\oplus}$ - $1000$\,$M_{\oplus}$) and orbital
distances ($0.0015$\,AU - $0.055$\,AU), searching for the masses that
produce an rms of $\sim30$\,s in the calculated central time of the
transits of GJ1214b during the ten years of integration time we used.
The results of these dynamical simulations are presented in
Fig.~\ref{fig:MvsA} where a region of unstable orbits is marked by the
gray strip. For all the other stable orbits we determined the
perturber mass that would produce a TTV rms of 30\,s (represented by
the solid line).  In the 1:2, 2:3 (interior), and in the 3:2, 2:1
(exterior) MMRs (vertical lines in Fig.~\ref{fig:MvsA}) the upper mass
limits we obtained correspond to $\sim0.5~M_{\oplus}$.  These are
better constraints in the mass of a possible companion of GJ\,1214\,b
than the limits imposed by radial velocity measurements.

\section{Summary\label{sec:summary}}

GJ\,1214\,b is undoubtedly one of the most intriguing objects and the
first of its class with an extensively studied atmosphere. The near-IR
and optical photometric measurements we present here provide
additional evidence for a rather flat featureless spectrum, indicating
either a metal-rich, or a cloudy or hazy atmosphere. The TTV analysis
of our data combined with previous data sets by other groups have
confirmed the constant value of the planetary orbital period.

All observations reported here were performed with 4m class telescopes
and prove the suitability of such facilities for high-precision
photometry. Furthermore, we encourage new spectroscopic measurements
especially in the $H$ and $K_{s}$ bands of the spectra. Particularly
suitable instruments are either multi-object spectrographs or, as
presented in this paper, long-slit spectrographs with very wide and
long slits capable of simultaneously monitoring a comparison star. In
the latter case, 4m class telescope usage as for GJ1214b may be
challenging, but might provide very interesting results for multiple
events.

\begin{acknowledgements}
  The authors are grateful to the SOAR and the ESO staff for the help
  during the observations. The SOI observations were performed thanks
  to granted Director's Discretionary Time of the SOAR Telescope. CC
  acknowledges the support from ALMA-CONICYT Fund through grant
  31100025 and project CONICYT FONDECYT Postdoctorado 3140592. SH and
  PR would like to acknowledge grant FONDECYT 1120299. S.H. also
  acknowledges financial support from the Spanish Ministry of Economy
  and Competitiveness (MINECO) under the 2011 Severo Ochoa Program
  MINECO SEV-2011-0187. PK acknowledges the co-funding under the Marie
  Curie Actions of the European Commission (FP7-COFUND) and MCMC code
  kindly provided by Mich\"{a}el Gillon (University of Li\`{e}ge). DM
  thanks Basal-CATA PFB-06.
\end{acknowledgements}

\bibliographystyle{aa}
\bibliography{submitted}
\end{document}